\documentclass[twocolumn,showpacs,pre,aps,superscriptaddress]{revtex4}
\usepackage[utf8]{inputenc}
\usepackage[english]{babel} 
\usepackage{graphicx} 
\usepackage{listings} 
\usepackage{amssymb} 
\usepackage{amsmath} 
\usepackage{longtable}
\usepackage{multirow} 
\usepackage{booktabs}
\usepackage{pst-all} 
\usepackage{pstricks-add}
\usepackage{units} 
\usepackage{listings} 
\usepackage{xcolor}
\usepackage{subfigure}

\usepackage{hyperref}

\usepackage{bbm}

\usepackage{natbib}

\begin{document} 

\title{{ Numerical Evidence for Approximate Consistency and Markovianity of some Quantum Histories in a Class  of Finite Closed Spin Systems}}

\author{Daniel Schmidtke}

\email{danischm@uos.de}

\affiliation{Fachbereich Physik, Universit\"at Osnabr\"uck,
             Barbarastrasse 7, D-49069 Osnabr\"uck, Germany}

   \author{Jochen Gemmer}

\email{jgemmer@uos.de}

\affiliation{Fachbereich Physik, Universit\"at Osnabr\"uck,
             Barbarastrasse 7, D-49069 Osnabr\"uck, Germany}

\begin{abstract}

{ 
Closed quantum systems obey the Schr\"odinger equation whereas nonequilibrium behavior of many systems
is
routinely described in terms 
of classical, Markovian stochastic processes. Evidently, there are fundamental differences between those two types of
behavior.
 We discuss the conditions under which the unitary dynamics may be mapped onto pertinent classical stochastic processes. This is first
principally addressed based on the 
notions of ``consistency'' and ``Markovianity.''  Numerical data are presented that show that the above conditions are
to good approximation 
fulfilled  for  Heisenberg-type spin models comprising 12-20 spins. The accuracy to which these conditions are met
increases with system size.}

\end{abstract}


\maketitle

\section{Introduction}
May quantum dynamics be mapped onto standard stochastic processes, especially in closed quantum systems? It is
widely agreed that the general answer to this question is no (even though there have been investigations in this 
direction \cite{nelson1,nelson2}). Since the mid-2000s there
has been increasing research activities in the field of
``equilibration'' and
``thermalization'' with respect to 
closed quantum systems, although the latter mechanisms are traditionally  associated with stochastic processes.  Most of
these research activities
have 
focused on the remarkable  fact that after some, possibly very long, time
\cite{irgendwas1,irgendwas2,irgendwas3}, the behavior of many observables
is very well be practically 
indistinguishable from standard phenomenological
equilibrium behavior, despite the fact that
the Schr\"odinger equation does not feature any attractive fixed point. Some of these
attempts follow
concepts of pure state quantum statistical mechanics \cite{Bendersky2008,Linden2009}, typicality
\cite{Goldstein2006,Goldstein2010,Reimann2007}, or eigenstate thermalization hypothesis \cite{Goldstein2010,Rigol2008}.
However, according to textbook-level physics, a multitude of systems not only reach equilibrium after an extremely
long time but also evolve towards it
in a (quick) way that
conforms with some master or Fokker-Planck equations. Moreover, there have recently been attempts to find and
explain
the emergence of Fokker-Planck-type dynamics in closed quantum systems \cite{NieSchmGem2013, lesanovsky1, lesanovsky2,
Cohon}. Here
we go a step
further in that direction and investigate
to what extent the quantum dynamics of certain observables in a specific system can be seen as being  in
accord not only with a Fokker-Planck equation, but also with the underlying stochastic process. The latter allows for
producing
individual stochastic trajectories. 
\\
The approach {presented here} is  based on two central notions: ``Markovianity'' and ``consistency''.\\
Despite Markovianity already being a somewhat ambiguous term with differing definitions in the context of open
quantum
systems,
cf. Refs. \cite{Rivas2014,Vacchini2011,Breuer2010}, we add below another definition which is furthermore
applicable
to closed quantum systems.
The definition is based on mathematical constructions which have already been used by Wigner \cite{Wigner1963}
to quantify  probabilities for the occurrence of subsequent events. Our notion of consistency is the same as the one
used in the context ``consistent histories'' which also deals with these mathematical constructions.
Very loosely speaking, it
quantifies the absence of 
coherence between different events. If the dynamics of some system with respect to
 some set of projectors is consistent, then the
evolution of the 
expectation values of those projectors is independent of whether those projectors are repeatedly measured in
time.

To avoid confusion it is important to note that (although measurements are mentioned) we neither address
an open system scenario nor
do we use open system analysis techniques \cite{Breuer2010}. This is to be contrasted with literature showing that in
open systems, like  the Caldeira-Leggett model, 
histories of, e.g., position measurements become consistent in the Markovian limit \cite{22,25,26}. While the
Caldeira-Leggett model is accessible by a 
Feynman-Vernon path integral approach that also allows for the formulation of consistency \cite{22,26}, our models
are not coupled to any baths, 
nor are their classical analogs integrable, thus rendering a path integral approach futile. A crucial point of our
investigation is precisely the fact 
that consistency and Markovianity may occur even without any kind of ``environment-induced superselection.''

The {present paper is organized as follows:  in Secs. \ref{sec:conditions} and \ref{sec:unitostoch} the 
``consistency concept'' is reviewed and our notion Markovianity is specified. Furthermore, we use the general point of
view that unitary dynamics may be 
mapped onto classical stochastic processes, if the unitary dynamics are consistent and Markovian, onto a more formal
basis. We also argue qualitatively that
typical Hamiltonians yield consistent and Markovian unitary dynamics for typical observables in Sec.
\ref{sec:qualitative}.\\
Section \ref{sec:numerics} contains 
our main result. It is a specific numerical example supporting the correctness of the qualitative argument given in the
previous Sec. \ref{sec:qualitative}. 
We numerically investigate some generic sequences of transitions (or ``quantum histories'') in a generic spin system. We
repeat the investigation for the same type of
spin system but for sizes of 12-20 spins. This finite-size scaling suggests that the addressed quantum histories become
indeed consistent and Markovian in the limit 
of large systems. Some comments on
many-step Markovianity are given in 
Sec. \ref{sec:morestep} where special attention is laid on
comparison of sequences of identical events to random event sequences. Conclusions are drawn and possible further
investigations are outlined in Sec. \ref{sec:conclusion}.}

\section{Consistency and Markovianity conditions}\label{sec:conditions}

Consistency is obviously a central concept in consistent history approaches \cite{Grifiths1984}, which is sometimes
also called decoherent histories \cite{22,Halliwell2009}. In the context of the {current paper} we only
need to
introduce what is sometimes called the 
``decoherence  functional'' or ``consistency condition'' and its properties. The more philosophic
aspects of the consistent history
approach, concerning the interpretation of 
quantum mechanics 
are of no relevance here (for critical reviews see, e.g., Refs. \cite{Allahverdyan2013,Okon2014}). However, we will
recapitulate
the basic notions of the consistent history 
approach, in order to enable the reader unacquainted
with the latter to develop a full understanding of the analysis within this paper.\\
Broadly speaking, one may describe quantum  histories as a method to deal with the occurrence probabilities of
certain event sequences. An event sequence consists of the alternating mathematical actions of measuring a certain
property, encoded in some set of 
projection operators $\pi_n$, according to the von Neumann measurement scheme, and time-propagating the resulting state
according to the Schroedigner equation. 
Note that the measured properties need not to be identical each time.\\
The term ``consistent`` expresses the accordance
of
history probabilities with the
Kolmogorov axioms, and the degree of accordance is quantified by the {previously mentioned} consistency
condition, which is
one of the central notions of the 
paper. With these preliminary remarks we embark on a
somewhat more formal 
presentation of the consistent history approach.\\

To begin with, we introduce a complete set of projectors, i.e.,
\begin{equation}
\sum_i \pi_i \,=\, \mathbbm{1}~~,
\label{eq:projectors}
\end{equation}
where $\mathbbm{1}$ denotes the unity operator and each projector corresponds to a possible event or measurement result.
The set of projectors corresponds to some property.

We denote by $\rho(t)$ the density operator describing the system at time $t$ and obtain the occurrence
probability of event $i$ at time point $t$ as
\begin{equation}
 P(x_i(t)) = \text{tr}\{\pi_i \rho(t)\}~~.
\end{equation}
To shorten the following expressions, we define two abbreviations, i.e., (i) $\Pi_i \,\rho = \pi_i \,\rho\,\pi_i$
and (ii) $\mathcal{U}(\tau)\, \rho = U(\tau)\, \rho\, U(\tau)^\dagger$, where $U$ denotes the time-translation operator
which
propagates the system states in amounts of $\tau$.\\
A history is now created by performing a time translation 
after each measurement. To each of these histories one now assigns an occurrence probability, e.g. the  event sequence
$x_1(0)\rightarrow x_2(\tau_1)\rightarrow  x_3( \tau_1+\tau_2 )$ 
{occurs} with the probability
\begin{equation}
\begin{split}
 P(x_1(0), x_2(\tau_1), x_3( \tau_1+\tau_2 ) ;\rho) =\\
 \text{tr}\{\Pi_3 \,\mathcal{U}(\tau_2) \,\Pi_1
\,\mathcal{U}(\tau_1)\, \Pi_1 \,\rho \}~.
\label{eq:probHist}
\end{split}
\end{equation} 
(This assignment was formally suggested in Ref. \cite{Wigner1963}, i.e., before the consistent history concept was
introduced.)

For simplicity, we will consider hereafter  only equal time steps, i.e., $\tau= \tau_1=\tau_2= \cdots$, and
therefore omit the time parameter.\\

{A special situation in the context of consistent histories arises if the observation is not continuous},
e.g., if one 
actually measures only in the beginning
and the end and leaves
the property at an intermediate time unmeasured. In this case (\ref{eq:probHist}) becomes
\begin{eqnarray}
 P(x_1,--, x_3;\rho) &=&  \text{tr}\{\Pi_k\, \mathcal{U}^2\,\Pi_i\, \rho \}\\
 &=& \sum_{x_2} P(x_1, x_2, x_3 ;\rho)\label{eq:probCon1}\\
 &+& \sum_{i \neq j} \text{tr}\{\Pi_3\, \mathcal{U}(\,\pi_{i}(\, \mathcal{U}\,
\Pi_1\, \rho)\,\pi_j)\}\label{eq:probCon2}
\end{eqnarray}
where $--$ indicates that at this intermediate time-point no measurement is performed.

The crucial expression here is Eq. (\ref{eq:probCon2}), since this is the above mentioned decoherence functional.
Only if the latter vanishes (which is called the consistency condition) does perfect accordance with the
third Kolmogorov axiom (KA 3) result.
Put another way: If some main event may be obtained as the result of many
different, independent ``subevents,'' then the 
probability for the main event to occur is given by the sum of the probabilities of the ``subevents.''\\
Since we will specifically calculate the value of decoherence
functional numerically for some concrete 
examples,  we rewrite the consistency condition here in an more explicit style,
\begin{equation}
\sum_{{i\neq j }} \text{tr}\{\pi_3\, U \,\pi_i\,U\, \pi_1\, \rho \,\pi_1\, U^\dagger\,
\pi_{j}\, U^\dagger\} \approx 0~~~.
\label{eq:consicon}
\end{equation}
It is not to be expected that this expression ever vanishes precisely in a generic situation based on a finite quantum
system, but it may possibly approach zero in the limit of infinitely 
large systems.
It is this latter statement which is one of the main targets of {this paper}. At this point we would like
to
emphasize some consequence of (\ref{eq:consicon}) 
for later
reference:    If (\ref{eq:consicon}) applies, then, by virtue of (\ref{eq:projectors}), summing the probability of some
quantum history over all possible events 
at specific times produces the probability of a quantum history in which there are no measurements at the  corresponding
times, e.g.:
\begin{equation}
\sum_{x_1, x_2} P(x_1, x_2, x_3 ;\rho) =  P(x_3 ;\rho) 
\label{eq:summing}
\end{equation}

Here we close our outline of  basic concepts in consistent histories and refer the interested reader to the pertinent
literature, e.g., Refs.
\cite{Grifiths1984,Omnes1987}, and turn towards Markovianity.\\

In the context of (open) quantum dynamics the term ``Markovian'' has been used for a variety of features
\cite{Breuer2010}. However, for the remainder of {
this work}, ``Markovianity'' will be used to describe a property of quantum histories. The rationale behind this
concept is that histories will be called 
Markovian, 
if a few past measurement outcomes suffice to fix the probabilities for the next future  measurement outcomes. Our
definition employs the notion of conditional 
probabilities as inferred from quantum histories. The construction of such conditional probabilities is straightforward,
and we simply define them as the
ratio of the occurrence 
probability
of the event sequence $x_k \rightarrow \cdots \rightarrow x_{n+1}$ to that of $x_k \rightarrow \cdots \rightarrow
x_{n}$, i.e.,
\begin{equation}
\omega(x_{n+1}|x_k, x_{k+1},\cdots, x_{n} ;\rho) = \frac{P( x_k, \cdots, x_{n}, x_{n+1};\rho)}{P( x_k, \cdots,
x_{n};\rho)}~~~.
\label{eq:conditional}
\end{equation}
We call such a conditional probability one-step Markovian if
\begin{equation}
\omega(x_{n+1}  | x_{n} ;\rho) = \omega(x_{n+1}|x_k, \cdots, x_{n};\rho)~~~
\label{eq:onestep}
\end{equation}
holds true, two-step Markovian if only 
\begin{equation}
\omega(x_{n+1}  | x_{n-1},  x_{n} ;\rho) = \omega(x_{n+1}|x_k, \cdots, x_{n};\rho)~~~
\label{eq:twostep}
\end{equation}
holds, and so on. Obviously, conditional probabilities $\omega$ themselves as well as the  validity of the above
equations
(\ref{eq:onestep} and \ref{eq:twostep}) depend on the initial state $\rho$. Below, in (\ref{rhoini}) we will focus on
a
specific class of initial states in 
order to get rid of this dependence.

\section{Qualitative consideration on the typicality of consistency and Markovianity}\label{sec:qualitative}

In the previous section consistency and Markovianity have been defined as properties of dynamics depending on both 
the Hamiltonian $\hat{H}$ of the system and the observable that is actually being watched, the latter being formalized
by the set of projectors $\{\hat{\pi}_i\}$.
Having a feasible scheme which allows us to decide whether, for given $\hat{H},\{\hat{\pi}_i\}$,  consistency and
Markovianity are present would be 
very instructive and generally most desirable. Unfortunately,  such a scheme is yet unknown (however, we consider its
development as an ambitious and
promising line of future research). Thus we primarily resort to numerics and give  in Sec. \ref{sec:numerics} a
concrete example for a system and an observable 
which is consistent and Markovian. 

Numerics, however, cannot answer the principal and important question 
if  consistency and Markovianity may, in some sense, be generally expected. While we are  far from being able to answer
the question conclusively, we outline  in 
the following a qualitative argument pointing in the direction of consistency and Markovianity being indeed natural for
systems and observables featuring large 
Hilbert spaces and few symmetries. The argument is along the lines of the more general concept of ``typicality''
\cite{Goldstein2006, Cho,SheldonGoldstein2010}.

Consider an {addend} of the sum which serves to specify consistency (\ref{eq:consicon}). Denote the eigenstates of the
projectors by
$\pi_i,\pi_{j}$ by $\{|n_i\rangle \}, \{|n_j\rangle \}$, respectively. Then a single {addend} for specific $i,j$ reads
\begin{equation}
\sum_{n_i, n_j}  \langle n_j|U^\dagger  \pi_3\, U \,|n_i\rangle  \langle n_i|  \,U\, \pi_1\, \rho \,\pi_1\,
U^\dagger\,|n_j\rangle
\label{eq:consum}
\end{equation}
For $i\neq j$ the above sum (\ref{eq:consum}) comprises products of two factors, both of which are complex numbers. The
phases of those numbers are neither
related to each other by any general principle nor restricted to a certain interval within the full range of
$]0,2\pi]$. Thus the terms in the
sum may `` average out'' to zero. Indeed, if $U$'s are drawn at random, {(such that the mapping of any pure state onto any other pure state is equally 
probable, cf. e.g., Ref. \cite{SheldonGoldstein2010})}, then the
averages over the individual factors vanish, 
as long as $\langle n_j|n_i\rangle = 0$ \cite{GemmerQT}. Furthermore, fluctuations around this average vanish as
$\propto
1/d$, where $d$ is the dimension of 
the respective Hilbert space \cite{GemmerQT}. Thus for the (overwhelming) majority of $U$'s  (\ref{eq:consum}) is
expected to result into a very small number, 
which then implies consistency. This is to be contrasted with the situation  $i= j$. In this case both factors of the {addends}
of (\ref{eq:consum}) are 
real, positive numbers. Hence summing many of them will typically yield a considerably larger positive number.

A similar argument can be formulated which indicates that Markovianity is typical in the same sense. Consider the
probability to get measurement outcome $x_3$ 
after $x_1$ and $x_2$ have occurred. According to (\ref{eq:conditional}) the corresponding conditional probability
reads 
\begin{eqnarray}
&&\omega(x_3|x_1, x_2;\rho) = \\ 
&&\frac{\sum_{n_2, m_2} \langle n_2|U^\dagger  \pi_3\, U \,|m_2\rangle  \langle m_2|  \,U\, \pi_1\, \rho \,\pi_1\,
U^\dagger\,|n_2\rangle  }
{\sum_{n_2} \langle n_2|  \,U\, \pi_1\, \rho \,\pi_1\, U^\dagger\,|n_2\rangle  } \nonumber
\label{eq:mark}
\end{eqnarray}
where $ |n_2\rangle,  |m_2\rangle $ are eigenstates of $\pi_2$. If one, based on the same argument as given below
(\ref{eq:consum}), drops all terms that are 
not necessarily real and positive, this reduces to:
\begin{eqnarray}
&&\omega(x_3|x_1, x_2;\rho) \label{eq:mark1}\approx\\ 
&&\frac{\sum_{n_2} \langle n_2|U^\dagger  \pi_3\, U \,|n_2\rangle  \langle n_2|  \,U\, \pi_1\, \rho \,\pi_1\,
U^\dagger\,|n_2\rangle  }
{\sum_{n_2} \langle n_2|  \,U\, \pi_1\, \rho \,\pi_1\, U^\dagger\,|n_2\rangle  } \nonumber
\end{eqnarray}
Again following the concepts of typicality one finds 
$ \langle n_2|  \,U\, \pi_1\, \rho \,\pi_1\, U^\dagger\,|n_2\rangle \approx 1/ (\text{tr}\{\pi_2\} \text{tr}\{\pi_1\,
\rho \,\pi_1\})   $ for the 
overwhelming majority of all randomly distributed) $U$. Inserting this into (\ref{eq:mark1}) yields:
\begin{eqnarray}
\omega(x_3|x_1, x_2;\rho) &\approx& \frac{\sum_{n_2} \langle n_2|U^\dagger  \pi_3\, U \,|n_2\rangle  }
{ \text{tr}\{\pi_2\}  } \label{eq:mark22} \\   
& =&\omega(x_3|x_2; \pi_2 )
\label{eq:mark2}
\end{eqnarray}
Thus, for the majority  of all  $U$, neither the concrete initial state $\rho$ nor the next-to-last observed value $x_1$
are relevant for the occurrence 
probability of $x_3$;  it is only the very last observed value  $x_2$ that matters. This is what has been defined as
one-step Markovianity in (\ref{eq:onestep}).
Hence, in this sense one-step Markovianity is typical. 

It should be emphasized here, that all the above reasoning is based on ``typical unitaries'' $U$. While such a
consideration is mathematically legitimate (and can 
be made rigorous \cite{Goldstein2006, Reimann2007}), it does not imply  that counterexamples do not exist. It does not
even necessarily imply  that counterexamples are 
rare in  nature. {Random} $U$'s are generated by Hamiltonians $H$ that are essentially random, Hermitian 
matrices. However, most 
quantum many-particle models are characterized 
by Hamiltonians that differ significantly from random matrices: They are often sparse with respect to the
site-occupation-number
basis, they usually have only 
real entries, etc. 
Hence the considerations presented in the current section by no means replace the concrete numerical computations in
Sect.
\ref{sec:numerics}.

\section{From unitary dynamics to one-step stochastic processes}\label{sec:unitostoch}

In this section, we establish that the dynamics of the above event probabilities, as following from the
Schrödinger equation for non measured closed systems, may be
rewritten as Markovian stochastic processes, provided that Eq. (\ref{eq:consicon}) and Eq. (\ref{eq:onestep}) hold. To
this end we
start by writing out the probability of some event $x_{n+1}$ at the corresponding time in a seemingly complicated
way, relying on (\ref{eq:summing}):
\begin{equation}
 \label{pnext}
 P(x_{n+1})= \sum_{x_1,\cdots, x_n}P(x_1, \cdots ,x_n, x_{n+1};\rho) 
\end{equation}
This shall be rewritten in an even more complicated fashion as:
\begin{equation}
 \label{pnextcond}
P(x_{n+1})= \sum_{x_1,\cdots, x_n}   \frac{P(x_1, \cdots ,x_n, x_{n+1};\rho)}{P(x_1, \cdots ,x_n ;\rho)} P(x_1, \cdots
,x_n ;\rho).
 \end{equation}
However, if one-step Markovianity holds, i.e., if Eq. (\ref{eq:onestep}) applies, the above fraction may be replaced by
the simpler one-step conditional probability,
\begin{equation}
 \label{pnextmark}
 P(x_{n+1})= \sum_{x_n} \omega(x_{n+1}  | x_{n} ;\rho)  \sum_{x_1,\cdots, x_{n-1}}P(x_1, \cdots ,x_n ;\rho). 
\end{equation}
Exploiting (\ref{eq:summing}) again, this can be written as
\begin{equation}
 \label{pnextmark}
 P(x_{n+1})= \sum_{x_n} \omega(x_{n+1}  | x_{n} ;\rho)  P(x_{n}). 
\end{equation}
Except for the dependence of the transition probabilities $\omega$ on the very initial state, this equation is
equivalent to a standard
definition  of a Markov chain on the sample space containing all $x$. For the remainder of this paper we specialize in
certain
initial states $\rho$ of the form
\begin{equation}
 \label{rhoini}
 \rho=\sum_i c_i \pi_i  \quad  c_i \geq 0~~.
\end{equation}
The motivation for this choice is twofold. First, it may be viewed as a state in accord with Jayne's
principle:
If nothing is known about a quantum state except for the probabilities $P_i$ of finding the outcome $x_i$, a state
$\rho$ of the form given in (\ref{rhoini})
with $c_i = P_i/\text{tr}\{\pi_i\}$ maximizes the von Neumann entropy 
subject to the information given. Second, (\ref{eq:mark2}) suggests that a state   
$\rho$ of the form of  (\ref{rhoini}) produces transition probabilities in accord with the Markovian transition
probabilities which are expected for typical 
unitaries $U$: It is simply a projector (in the specific example $\pi_2$) that takes the role of the initial state in 
(\ref{eq:mark2}).

From Eq. (\ref{eq:conditional}) it may also be inferred that for this class of initial states the transition
probabilities
$\omega$ are actually independent of the actual $c_i$, i.e., the 
transition probabilities $\omega$ are all the same for the entire class of initial states. Due to this, we omit, for
brevity, $\rho$ in the argument of 
 $\omega$, thus obtaining 
\begin{equation}
 \label{pnextmark}
 P(x_{n+1})= \sum_{x_n} \omega(x_{n+1}  | x_{n} )  P(x_{n}). 
\end{equation}
which defines a standard Markovian stochastic process.

To recapitulate the analysis so far, it can be stated that, if consistency and one-step Markovianity hold, it is
straightforward to
demonstrate that the unitary time propagation according to Schrödinger equation can also be expressed by a
time-discrete stochastic process. Obviously the fact that we used one-step Markovianity is not necessary, i.e., also for
more-step Markovianity stochastic processes may be formulated in an analogous way. However, since the models we
investigate below appear to exhibit one-step
Markovianity to sufficient accuracy, {we will restrict ourselves to this case in the present section.}\\

For convenience, at this point  we do not (re-)define the features consistency and Markovianity directly but rather
quantify their complements  ``nonconsistency'' $\bar{C}$   and 
``non-Markovianity'' $\bar{M}$. Both are below defined to be real numbers with $0\leq \bar{C}(\bar{M})$ in such a way
that 0 indicates perfect consistency 
(Markovianity) and any larger value expresses a (gradual) violation of the respective feature. {The
definition of
nonconsistency reads}:
\begin{equation}
\bar{C}(x_k,  \cdots, --, \cdots, x_n) = \bigg \vert 1-\frac{P(x_k,  \cdots, --, \cdots, x_n)}{\sum_\gamma P(x_k,
\cdots, x_n  )} \bigg
\vert~~~,
\label{eq:noncon}
\end{equation}
{where summation over all possible intermediate outcome sequences $\gamma$ is     
denoted here as $\sum_\gamma$. In a similar fashion non-Markovianity is defined by}:
\begin{equation}
\bar{M}(x_k, \cdots, x_n) = \bigg \vert 1-\frac{\omega(x_{n+1}|  x_{k-1}, \cdots, x_n)}{ \omega(x_{n+1}|x_{k}, \cdots,
x_n)}\bigg \vert~~~,
\label{eq:nonmark}
\end{equation}
It may be worth noting here that this is not the only possible sensible definition of non-Markovianity even within this
framework. It obviously refers only 
to some specific conditional probability and  takes only one prior measurement into account. In the remainder of this
paper we will mainly focus on the exemplary
investigation of some specific measurement outcome sequences, and return to more general questions in Sec.
\ref{sec:morestep}.

\section{Numerical investigations}\label{sec:numerics}

\begin{figure}[h]
 \centering
\includegraphics[scale=0.225]{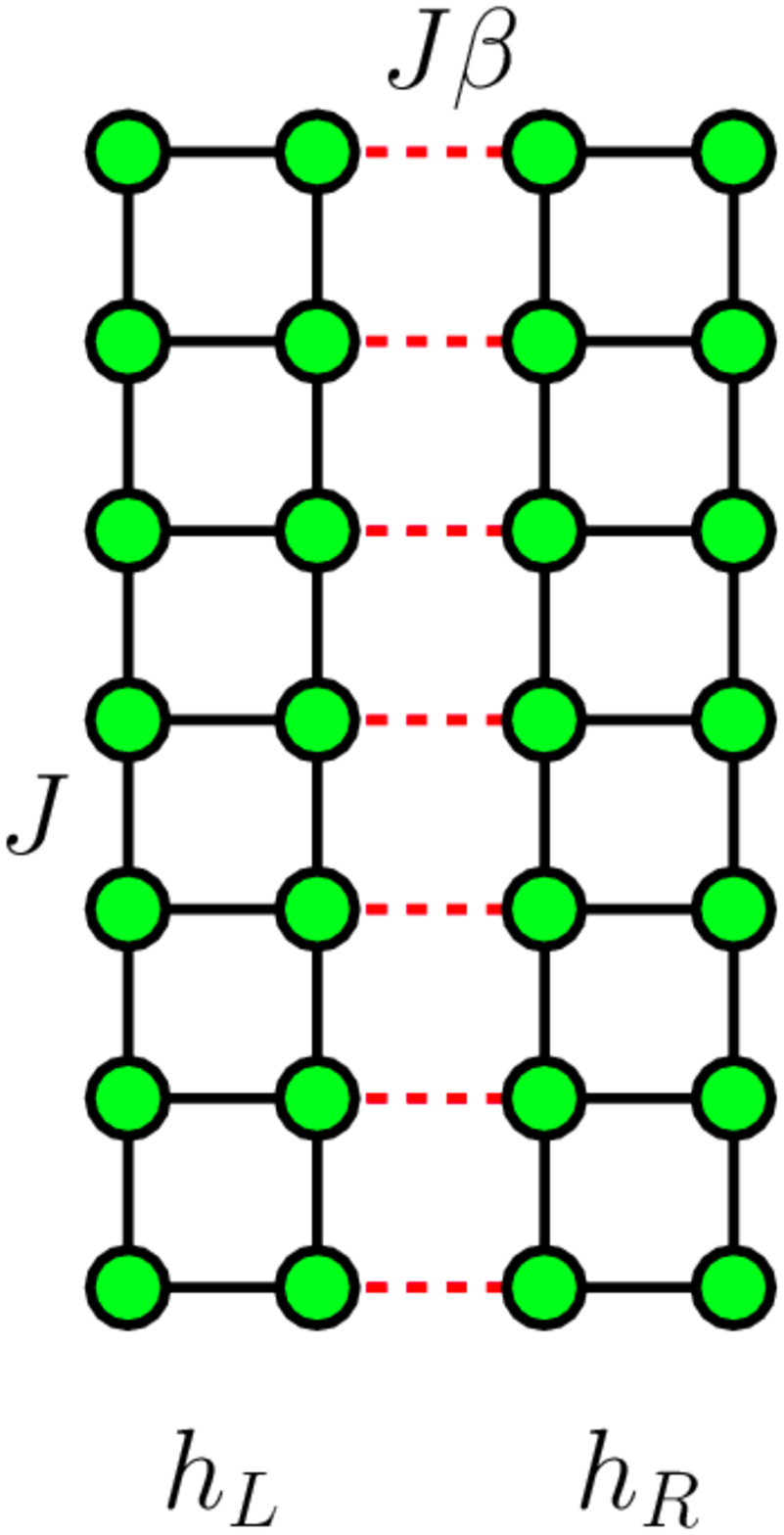}
  \caption{(Color online) Schematic visualization of a spin lattice consisting of two separable spin ladders. Each spin
ladder has an
assigned Hamiltonian $h_L$ ($h_R$) where the spins are coupled by a strength $J$ (solid lines). Both ladders are in turn
coupled by a strength of order $J\,\beta$ (dashed lines).}
  \label{pic:model}
\end{figure}

The spin lattice that we are going to investigate basically consists of two spin ladders,{ with total
number of spins $N
= 4 n$}, which are brought into contact
along opposing spines, cf. model in Fig. \ref{pic:model}. Hence the Hamiltonian consists of three parts:
\begin{equation}
H = h_L \otimes h_R + V~~,
\label{eq:hamLat}
\end{equation}
where $h_L ~(h_R)$ denotes the local Hamiltonian of the left (right) subsystem. $V$ comprises the 
interaction between the subsystems.
The left local  Hamiltonian is defined by
\begin{eqnarray}
h_L &=& J \sum^n_{i=1} s^x_i s^x_{i + 1} +s^y_i s^y_{i + 1}+\Delta s^z_i s^z_{i + 1}\notag\\
&+&J  \sum^n_{i=1}
s^x_i s^x_{i + n} +s^y_i s^y_{i + n}+\Delta s^z_i s^z_{i + n} \notag\\
&+& \text{ h.c}.
\label{eq:hamLatLeft}
\end{eqnarray}
where the $s_{\cdots}^{\cdots}$ denote the pertinent operators of components of $s=1/2$ spins sitting at the respective
positions.
The  Hamiltonian of the right subsystem, $h_R$, is obtained through
shifting the indices in (\ref{eq:hamLatLeft}) by $2n$, i.e., $s_i \rightarrow s_{i+2n}$. The overall energy scale is
set by $J$.\\
The interaction of the both subsystems takes place only between the two ``central'' chains of the lattice, namely in
this model between the second and third chain. Thus the interaction term reads
\begin{equation}
V = J \beta \sum^{2n}_{i=n+1} s^x_i s^x_{i + n} +s^y_i s^y_{i + n}+\Delta s^z_i s^z_{i + n} + \text{ h.c}.
\label{eq:hamLatInt}
\end{equation}

The observable (or property) we are going to analyze in detail is the  magnetization difference between both subsystems,
i.e.,
\begin{equation}
X = \sum s^z_{i,L} - \sum s^z_{i,R}~~~,
\label{eq:magnetdiff}
\end{equation}
where each sum represents the present total magnetization in the $z$ direction within the left (right) spin ladder.
Furthermore we restrict our analysis to the 
subspace of vanishing total magnetization, i.e., $\sum s^z_{i,L} + \sum s^z_{i,R}=0$. Note that the latter is a constant
of motion in this model.  This subspace was 
essentially chosen since it is the largest one with respect to the dimension of the corresponding Hilbert space. 
Furthermore, we choose our event-operators corresponding to the property $x$ concretely as the following projectors: 
\begin{equation}
\pi_{ x, E} = \pi_E \, \pi_{x}\, \pi_E~~,
\label{eq:initDensisty}
\end{equation}
where $\pi_{x}$ is the projector spanned by all eigenstates of $X$ featuring the same eigenvalue $x$, i.e., $X=\sum_x x
\pi_{x}$. The projector $\pi_E$  restricts 
the dynamics to a more or less narrow region in energy space: It is spanned by all energy eigenstates of the uncoupled
system, (i.e., without taking $V$ into 
account) that feature eigenvalues with $E_i \in [-1.2 J \,, +0.6J]$. To put this another way: The full energetic width
of the system is on the order of the  number
of of spins, i.e., $NJ$. Moreover the chosen interval contains the highest densities of states with respect to energy.\\
Note that
since
$[\pi_E ,\pi_{x}]=0$ the $\pi_{ x, E}$ are in fact 
orthogonal projectors. Obviously, the $\pi_{ x, E}$ are not complete in 
the sense of (\ref{eq:projectors}). However, a formally complete set may always be introduced by adding the complement 
$\bar{\pi} = \mathbbm{1} - \sum_x \pi_{ x, E}$ to the $\pi_{ x, E}$'s themselves. Practically, this hardly makes any
difference since our numerics confirm that 
for our below choices of the model parameters almost no probability ever goes to  $\bar{\pi}$, i.e., $P( \bar{\pi}, t) <
10^{-4}$.

\begin{figure}[h]
 \centering
\includegraphics[scale=0.55]{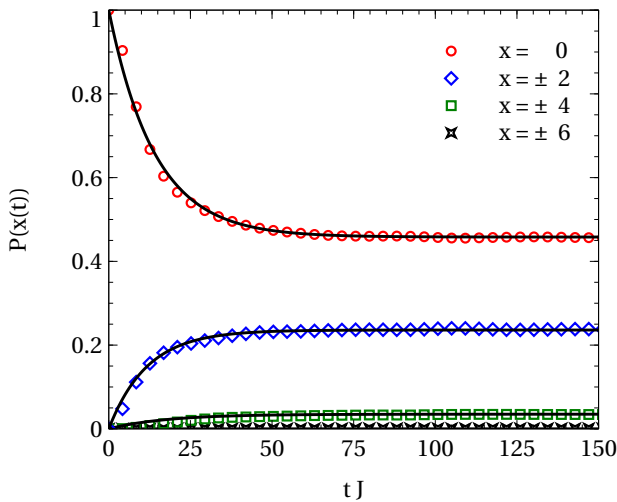}
 \caption{(Color online) Displayed are the dynamics of $P[x(t)]$ starting from $P(x=0)=1$ at $t=0$ for all occurring
values of $x$
within a system of size $N = 12$, i.e., $x \in \{0, \pm 2, \pm 4 , \pm 6\}$. These dynamics are compared to results from
simple transition-rate-based master equations (solid lines). Agreement indicates that a fully stochastic description
may be possible.}
 \label{rough}
\end{figure}

Before we turn towards numerical results on consistency and Markovianity, we should point out that the whole setup,
i.e., the Hamiltonian, the observable, the 
energy shell, etc., have been chosen in the specific way described above in order to find a nonrandom, finite system,
in which consistency and Markovianity  emerge already for rather small systems.\\
There are results in the literature
that
suggest to those ends a set-up like the one defined above:
In Refs. \cite{NieSchmGem2013, lesanovsky1, lesanovsky2, Cohon} Fokker-Planck-type dynamics have been reported for more
or
less similar spin
systems. Furthermore, results in Ref. \cite{Khodja}
indicate that the so-called eigenstate thermalization hypotheses (ETH) may be best fulfilled for bipartite systems in
which the local subsystems are not merely 
spin chains. (Since the ETH guarantees a single, attractive, long-time probability distribution of the events its
applicability is necessary for the emergence of 
effective stationary stochastic process dynamics.) For a first rough and of course non-sufficient check of whether the
dynamics of our model may be in 
accord with a stochastic description, we compute the dynamics of the $P[x(t)]$, starting from $P(x=0)=1$ at $t=0$. The
result
is
displayed in Fig. \ref{rough}. 
The solid lines are 
obtained by solving simple transition-rate-based master equations. The agreement indicates that a fully stochastic
description may be possible. Of course, since the model is finite, there will be (quasi-) recurrences.
However, these are expected at 
times that are by magnitudes larger than any timescale considered here and thus excluded from our analysis. We fix the
principal time scale of interest  
by means of Fig. \ref{rough}. Although the true dynamics are strictly unitary, the  $P[x(t)]$ appear to relax to towards
constant values. Thus we call the 
the time scale at which this relaxation happens the ``relaxation time'' $\tau_R$. Specifically, we infer $\tau_R=20J$.

\begin{figure}[h]
 \centering
 \includegraphics[scale=0.55]{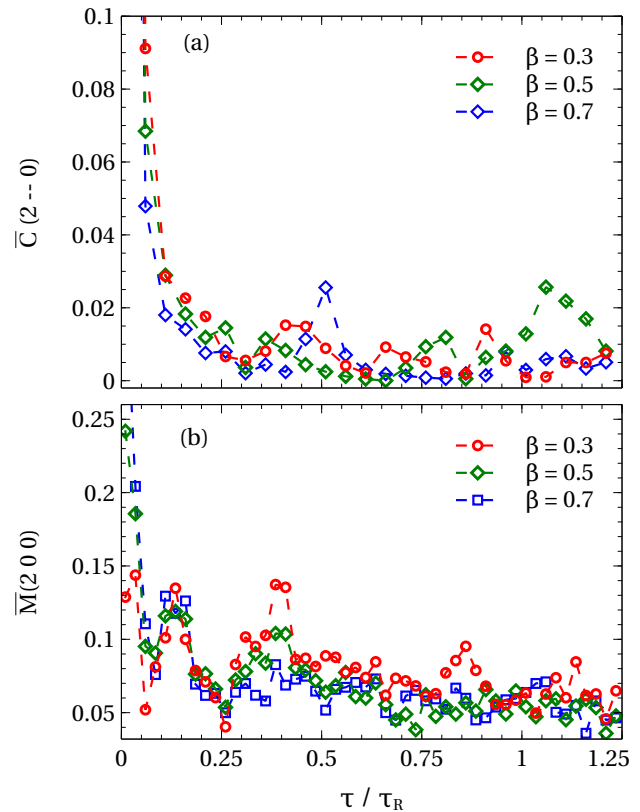}
 \caption{(Color online) Specific non-consistencies  (a) and (one-step) non-Markovianities  (b) are
displayed for
three different interaction strengths, 
depending 
 on the ``waiting time'' $\tau$ between measurements. The latter are given in units of the respective relaxation time
$\tau_R$. The system size is fix at $N=12$. 
 For waiting times $\tau $ larger than, say, $\tau_R/10$ both {nonconsistency and non-Markovianity,
remain, while fluctuating, low compared to unity} 
 at
all interaction strengths. 
 For our further exemplary investigations 
 we thus choose $\beta=0.5$
and $\tau = 0.5 \tau_R$.}
 \label{fig:nonconsistency1}
\end{figure}

We now investigate nonconsistency and non-Markovianity as defined by (\ref{eq:noncon}) and (\ref{eq:nonmark}),
in more detail. {As already pointed out before (\ref{eq:initDensisty}), nonconsistency and non-Markovianity here
refer 
to sequences of transitions between certain
magnetization differences. More precisely: the projectors which enter the definition of $\bar{C}, \bar{M}$ through
(\ref{eq:probHist}) are 
the projectors $\pi_{ x, E}$ as appearing in (\ref{eq:initDensisty}). Hence, below $x$ continues to indicate the 
magnetization difference.}
To begin with we kept the number of spins fixed at  $N=12$ and calculated $\bar{C}, \bar{M}$ as functions of $\tau$ (the
time elapsed between measurements)
for various coupling strengths $\beta$.  The paths chosen
for this example are $ x = \,2 \rightarrow -- \rightarrow 0$ (nonconsistency) and $x = 2
\rightarrow 0 \rightarrow 0$ (non-Markovianity). The results are displayed in  Fig. \ref{fig:nonconsistency1}.

\begin{figure}[h]
 \centering
\includegraphics[scale=0.55]{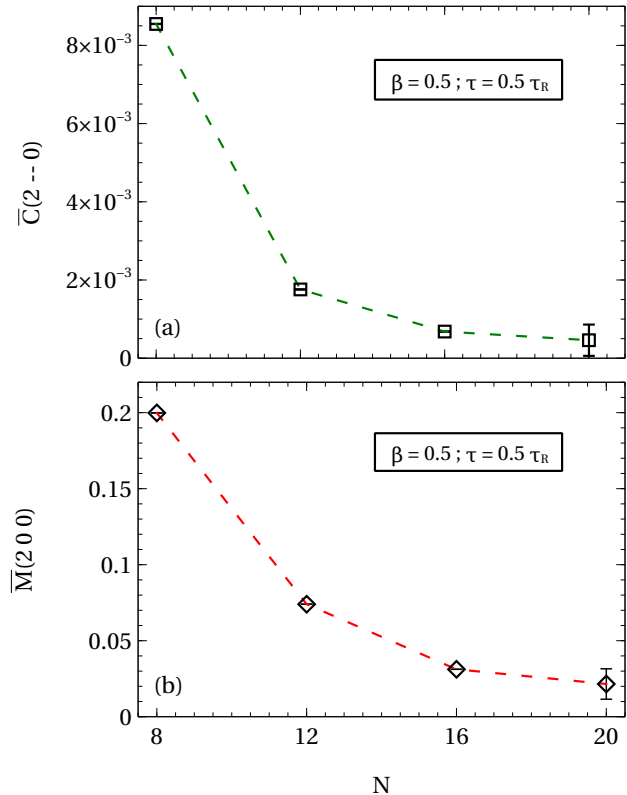}
 \caption{(Color online) Specific non-consistencies  (a) and (one-step) non-Markovianities  (b) are
displayed for 
 increasing system sizes $N$. Obviously both decrease with increasing system size. Although the data are not conclusive,
it suggests that both may vanish 
 in the limit of $N \rightarrow \infty$.}
 \label{fig:nonMarkovianity2}
\end{figure}

Though the graphs exhibit  rather large values for small portions of relaxation time, they decrease significantly at
times on the order of 
a tenth of the total relaxation time, i.e., $\tau \approx 0.1\, \tau_R$. Qualitatively, this behavior is the same for
all
investigated interaction strengths.
This indicates that there is a lower limit on the time step $\tau$, below which 
neither consistency nor Markovianity may be expected. This limit may, however, depending on the size and the structure
of the system, only be a small fraction of 
the total relaxation time. Precisely finding the minimum time step which allows for consistency and Markovianity is left
for
further research.
In the this paper we focus on a relatively large time step, i.e., $\tau = 0.5 \tau_R$, and primarily investigate
the
effect of increasing 
system sizes. Furthermore, we restrict our further analysis to interaction strength $\beta = 0.5$. The result (which is
our main numerical result) is displayed in 
Fig. \ref{fig:nonMarkovianity2}. It shows nonconsistency and non-Markovianity for various system sizes $N$. Up to
$N=16$ the results have been computed by means 
of direct numerical diagonalization. Due to limitations in computing power, we computed the result for $N=20$ using a
numerical method based on dynamical 
typicality. This method has been used and described e.g., in Refs. \cite{Steinigeweg2014, Bartsch2009,
NieMiDRaeGem2014}.
Based on this method we are able to address $N=20$ within 
reasonable computing time; however, the method involves random numbers and is thus subject to statistical errors. The
magnitude of the latter is indicated by the 
corresponding error bars. Obviously nonconsistency and non-Markovianity are  already small for moderate system sizes.
Furthermore, both decrease monotonically 
with increasing system sizes. {Figure \ref{fig:nonMarkovianity2} suggests that the dynamics 
become consistent and 
one-step Markovian in the limit of infinitely large systems. Whether this is indeed the case is
not to be answered conclusively from our finite-size scaling. It is possible to perform the same numerical calculations
for system sizes
up to, say,
$N=36$ \cite{NieMiDRaeGem2014}, but this requires high performance computing clusters. The present analysis, however, has been done using standard desktop 
computing equipment.}

\section{''More-than-one''-step Markovianity}\label{sec:morestep}

So far we primarily focused on one-step Markovianity throughout this paper. Furthermore, the numerical analysis in the
previous section was based on a specific 
definition 
of one-step Markovianity (\ref{eq:nonmark}) that takes only one prior outcome into account. Such an analysis is
necessarily  insufficient for the rigorous mapping of unitary dynamics onto a stochastic process. This may be seen most
easily 
from considering two aspects: (i) If a stochastic process is not fully Markovian with respect to one-step Markovianity,
it may
nevertheless be possibly fully
Markovian with respect to, e.g.,  two-step Markovianity. Thus some  finite one-step non-Markovianity does not rule out a
process from being Markovian altogether.
(ii) Even if it is found that the conditional probability to get some event at time $(n+1)\tau$ does not change much if
one takes the measurement outcome not 
only at time $n\tau$ but also at  time $(n-1)\tau$ into account, this does not \textit{a priori} mean that the
conditional
probability does not change much 
if, e.g.,  the  outcome at  time $(n-2)\tau$ is additionally taken into account. However, as explained below
(\ref{eq:nonmark}) only the former feature is 
captured by the definition  of non-Markovianity and numerically analyzed in Sec.
\ref{sec:numerics}. We emphasize again that the concept of ``repeated measurements'' employed here does
not evolve any 
outside measurement apparatus or any external environment. It exclusively refers to histories as formulated in
(\ref{eq:probHist}) and is thus well defined 
also for closed systems. Hence, the present consideration is not to be confounded with the Zeno effect where quickly
repeated external measurements ``freeze''
the dynamics \cite{kupschjooszehkiefer}. While being always  well defined our concept primarily addresses repeated
measurements with time steps larger than the time
at which consistency vanishes in the short time limit, cf. Fig. \ref{fig:nonconsistency1}.

A full-fledged numerical analysis taking all possible histories  and all above aspects of many-step-Markovianity
exhaustively into account is  beyond our 
possibilities, given the limit of reasonable computing time. However, in the following we focus on the many-step
Markovianity of some special histories.

The first history we address is the one that is generated by getting, upon repeated measuring, always the same outcome.
A
history like this may be relevant in 
situations in which some measurement outcome corresponds to the equilibrium state of the system. As will be explained
below, it turns out that such a history
is necessarily $\lambda$-step Markovian,  in the limit of large $\lambda$, irrespective of the concretely  considered
system.

Considering the occurrence probabilities for this type of history with initial states of the class introduced in
(\ref{rhoini}) yields $P(x_i(0), x_i(\tau),\cdots ,x_i(\lambda \tau)) := \text{tr}\{ \pi_{i} U \pi_{i}U
\cdots U^\dagger  \pi_{i} U^\dagger\pi_{i} \}$, where $x_i$ characterizes the measured property and the second index
$\lambda$ labels the number of
performed measurements. For brevity the index $i$ will be omitted 
hereafter.
Let us denote the  eigensystem of the (non-Hermitian) matrix $U^\dagger \pi$ by  $U^\dagger \pi
|\varphi_n>=\phi_n
|\varphi_n>$, and the occurrence probability of $\lambda$ identical
measurements by $P(\{x\}_\lambda)$. Then this  occurrence-probability is given by
\begin{equation}
P(\{x\}_\lambda) = \sum_{ijk} c^*_{ik} c_{jk} \{\phi^*_i\}^\lambda \{\phi_j\}^\lambda~~~,
\label{eq:omegaN}
\end{equation}
where the $c_{ij}$ denote the complex matrix element of the transformation that maps the nonorthogonal eigenvectors
$|\varphi_n>$ onto an orthonormal basis, 
i.e., $\sum_{ij} <\varphi_i|\varphi_j>  c^*_{ik} c_{jl} =\delta_{kl}$ .       

All eigenvalues of $U^\dagger \pi$ are upper bounded by 1, hence $|\phi_i| \leq 1$ holds, and thus
$\{\phi^\lambda_i\}$
describes a convex sequence with respect to
$\lambda$. Consequently, $P(\{x\}_\lambda)$, consisting only of sums of convex functions, is also convex. 
Convexity implies 
\begin{equation}
\frac{P(\{x\}_{\lambda+1})}{P(\{x\}_{\lambda})} \leq \frac{P(\{x\}_{\lambda+2})}{P(\{x\}_{\lambda+1})}~\forall~\lambda
\label{eq:convex}
\end{equation}
hence $\omega_\lambda$, defined as 
\begin{equation}
\omega_\lambda=\frac{P(\{x\}_{\lambda+1})}{P(\{x\}_{\lambda})}
\label{eq:omegan}
\end{equation}
is a monotonously increasing sequence. From the definition of the history probability we immediately find 
$P(\{x\}_{\lambda+1}) \leq P(\{x\}_{\lambda})~~\forall \lambda$. Thus  $\omega_\lambda$ is upper bounded by one. Since
$\omega_\lambda$ is monotonously increasing
but upper bounded it must converge against some finite value $d\leq 1$:
\begin{equation}
\lim_{\lambda \rightarrow \infty} \omega_\lambda    = d~\in~\mathbb{R}^+.
\label{eq:convex2}
\end{equation}
Plugging this result  into the definition of non-Markovianity (\ref{eq:nonmark})     yields
\begin{equation}
\lim_{\lambda\rightarrow \infty} \bar{M}_\lambda=\bigg \vert 1 - \lim_{\lambda\rightarrow \infty} 
\frac{\omega_{\lambda+1}
}{\omega_\lambda} \bigg \vert = \bigg \vert 1 - \frac{d}{d} \bigg \vert \equiv 0~~~.
\label{eq:convexMark2}
\end{equation}
Hence, in case of repeatedly measuring some property and {consequently} obtaining identical events as
measurement
outcomes, perfect
Markovianity always results for
sufficiently many steps. Although the implications of this result are limited (it only applies to a single type of
history and takes consistency for granted)
we consider it a valuable point of reference.
\begin{figure}[h]
 \centering
\includegraphics[scale=0.55]{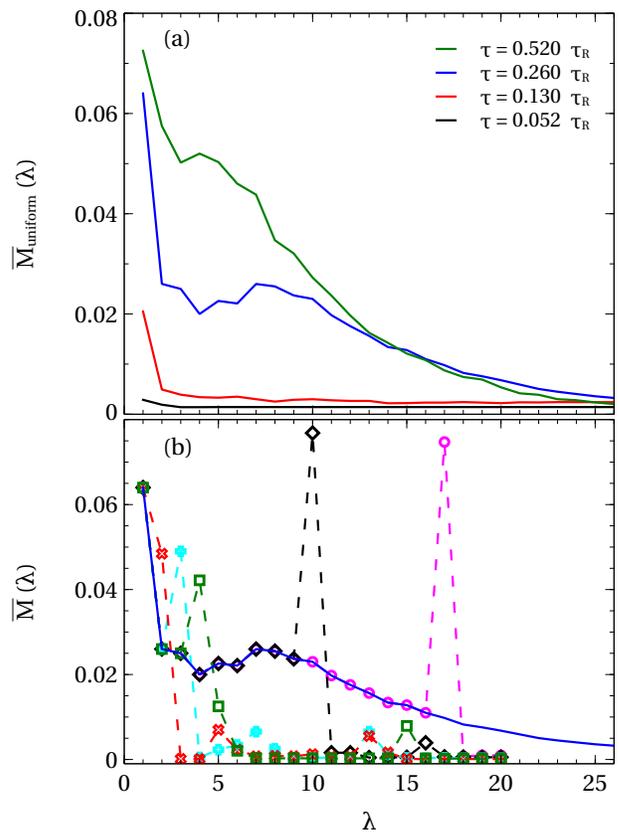}
 \caption{(Color online) Non-Markovianity for sequences comprising $\lambda$ measurements which yield all $x=0$ (a) or generated from 
 drawing random numbers according to the one-step transition probabilities $\omega$ (b). Data in (a)
confirm that 
 $\lambda$-step  transition probabilities always become perfectly Markovian in the limit of $\lambda \rightarrow \infty$
as claimed in the text.
 This result is independent of the waiting time $\tau$. Data in (b) suggest  that 
 $\lambda$-step  transition probabilities may  become perfectly Markovian in the limit of $\lambda \rightarrow \infty$
also in the case of 
 ``generic stochastic trajectories.'' However, at the first deviation of the measurement sequence from a uniform
sequence, somewhat larger 
 non-Markovianities occur; for more comments see text.}
 \label{fig:morestep}
\end{figure}

The data displayed in Fig. \ref{fig:morestep} address $\lambda$-step non-Markovianity for both histories of the
previously discussed type featuring identical 
outcomes, $x=0$, in Fig. \ref{fig:morestep}(a) and  for  some  random event sequences of the typical type which would
occur if one
simply took $\omega(x_{n+1}  | x_{n} )$ as a fully
one-step Markovian transition probability in Fig. \ref{fig:morestep}(b). Since jump probabilities away from $x=0$ are
low but towards
$x=0$ are high, the latter typical random 
sequences are, similarly to the former,  characterized by measuring $x=0$ most of the times, but exhibit
occasional ``excursions'' towards $x\neq 0$. We display 
data for $N=12$ and various $\tau$ in Fig. \ref{fig:morestep}(a) and  $\tau = 0.26 \tau_R$ in (b).

The graphs in (a) are obviously in accord with (\ref{eq:convexMark2}): Regardless of the ``waiting-time'' 
$\tau$, histories become Markovian in the 
limit of large $\lambda$. Furthermore, non-Markovianity is not strictly monotonously decreasing with $\lambda$, but in
the addressed data sample 
$ \bar{M}_{\lambda > 1} <\bar{M}_{\lambda = 1}$ appears to be strictly obeyed. Furthermore sequences appear to be more
Markovian for shorter waiting times.
Altogether one may conclude that for the uniform histories Markovianity appears to improve if more steps are taken into
account but a restriction to the 
``one-step-level'' may nevertheless be a very reasonable approximation. 

Considering the random histories in (b) it should first be noted that, while $ \bar{M}_{\lambda > 1}
<\bar{M}_{\lambda = 1}$ 
no longer strictly holds, non-Markovianities nevertheless remain very moderate also on the many-step level. Thus, also
in these cases a restriction to the 
one-step level appears to be a very reasonable approximation. However, the peaks towards relatively higher
non-Markovianities always occur at 
the most recent deviation from measuring identical outcomes. For example, in the history represented by the magenta
circles, the first (past)
16 outcomes are $x=0$, but the 17'th outcome is $x=2$. Nevertheless, in all our examples, while Markovianity becomes 
worse
for this most recent  deviation, 
it becomes better again with taking even longer histories into account. Thus, considering Fig.
\ref{fig:morestep}(b) one may guess that a statement 
like (\ref{eq:convexMark2}) also holds for arbitrary histories, and whether or not this holds true remains a subject of
future research.

\section{Summary and conclusion}\label{sec:conclusion}

{The possibility of describing the unitary dynamics as generated by the Schr\"odinger
equation quantitatively in terms of 
pertinent stochastic processes is addressed}. We discuss this possibility based on the notions of ``consistency'' and
``Markovianity.'' While the former refers to the concept of 
consistent histories, the latter denotes the independence of probabilities for  future measurement outcomes from
measurement outcomes in the distant past. We outline how a mapping mapping of quantum onto stochastic
dynamics can be performed  if, indeed,  the quantum dynamics is both consistent and
Markovian. This obviously leads directly to the question whether closed system dynamics are approximately
Markovian and consistent for specific, finite 
closed systems. This question is exemplarily discussed in the remainder of the paper.
The degree to which the 
quantum dynamics are indeed consistent and Markovian is specified by introducing corresponding quantifiers. These
quantifiers are 
numerically evaluated for a specific type of spin system. { By means of finite-size scaling we give
(strong) evidence
that the dynamics of this closed spin system can
indeed be considered consistent and 
Markovian.} A somewhat more detailed analysis shows that one may rely on a description based on stochastic processes
that
take only the most recent
past event into account. While an exhaustive numerical check of ``consistency'' and ``Markovianity,'' covering all
aspects of ``stochasticity,'' is far beyond  
of what can be done in finite computing time, our results indicate that a dynamical stochastic description of closed
quantum systems may be justified, even for 
rather small systems.

\section{Acknowledgements}

J. Gemmer acknowledges valuable input received at meetings 
of the   European network COST Action MP1209 “Thermodynamics in the Quantum Regime”.

\bibliography{references1}

\end{document}